\documentstyle[aps,multicol,epsf]{revtex}

\begin{document}
\draft

\title{
Effect of the accelerating growth of 
communications 
networks on their structure 
}

\author{
S.N. Dorogovtsev$^{1, 2, \ast
}$ and J.F.F. Mendes$^{1,\dagger}$
}

\address{
$^{1}$ Departamento de F\'\i sica and Centro de F\'\i sica do Porto, Faculdade 
de Ci\^encias, 
Universidade do Porto\\
Rua do Campo Alegre 687, 4169-007 Porto, Portugal\\
$^{2}$ A.F. Ioffe Physico-Technical Institute, 194021 St. Petersburg, Russia 
}

\maketitle

\begin{abstract}
Motivated by data on the evolution of the Internet and World Wide Web we 
consider scenarios of self-organization of the nonlinearly growing networks into free-scale structures. We 
find 
that the accelerating growth of the networks establishes their structure. 
For the growing networks with preferential linking and increasing density of links, two scenarios are possible. In one of them, the value of the exponent $\gamma$ of the connectivity distribution 
is between $3/2$ and $2$. 
In the other, $\gamma>2$ and the distribution is necessarily non-stationary.  
\end{abstract}


\begin{multicols}{2}

\narrowtext


The incredible place of the Internet in our civilization evokes the exponentially increasing 
flow of the studies of evolving networks (Internet, World Wide Web, neural networks, nets of citations, collaboration networks, etc.) 
\cite{hppj98,ajb99,ha99,ls98,r98,ba99,ba00a,asbs00,mn00}. The study of communications networks has a long history (e.g., see the papers of ``the father'' of the Internet, P. Baran \cite{b1}). Nevertheless, the first data on the structure of the Internet, Web, etc
were obtained only recently \cite{hppj98,ajb99,ha99,krrt99,fff99,bkm00}. Most of them are on the simplest ``one-site'' characteristic of networks -- the connectivity distribution. The observation of 
power-law 
dependences of the connectivity distribution in these networks 
puts forward the problem: 
how a growing network can organize itself into a scale-free structure? 

The most natural mechanism of such self-organization is the network growth with  
the preferential 
attachment of new links to sites with a high number of connections \cite{ba99,krrt99,baj99}. There are several ways to introduce the preferential linking which lead to different values of scaling exponents of the connectivity distribution inside of a huge interval \cite{dms00,krl00,dm00}. 
Nevertheless, it is still unclear why the exponents of the real systems have their specific values, and why the connectivity distributions have their specific forms? 
E.g., the last measured value for the scaling exponent of the distribution of a number of incoming links in the World Wide Web is $2.1$ \cite{ajb99,krrt99,bkm00}. 

The last data on the Internet evolution demonstrate that the total number of links increases more quickly than the number of sites \cite{fff99}. 
For instance, the study of the inter-domain topology of the Internet \cite{fff99} has shown that the relation of the edges and the nodes equaled $1.71$ in November of 1997 
(3015 nodes) and $1.88$ in December of 1998 (8256 nodes). The number of pages in the World Wide Web was $203\times 10^6$ with $1466\times 10^6$ links between them in May 1999 (the corresponding ratio is $7.22$) and it was already $271\times 10^6$ pages and $2130\times 10^6$ links in October 1999 (the ratio is $7.86$) 
\cite{bkm00}. 
That looks quite naturally -- old sites may establish new connections all the time. 
Old documents of the Web may be updated and new references may be added to them. 
Collaboration networks becomes to be more dense while growing because of increasing possibility to find a collaborator. 
Thus we face certainly the wide spreaded situation: the density of the links of a network becomes to be more and more high during its evolution, the total number of links grows more fast than the number of sites, the growth is nonlinear, that is {\em accelerating}. 

This very significant factor is omitted in most of considered models \cite{ba99,krrt99,baj99,dms00,krl00,dm00}. One may ask: What effect has the accelerating growth on the structure of the networks? 
What kind of the acceleration can produce scale-free networks? What form can take the connectivity distribution in such networks? 

In the present Communication we consider these problems using general arguments. 
Assuming that the networks are scale-free, we describe the possible connectivity distributions and show that, in this case, the total number of links is a power-law function of the network size. We introduce simple models with preferential linking accounting for the accelerated growth and study their evolution both analytically using the continuous connectivity approach 
\cite{baj99,dms00,dm00} and by simulation. The arising distributions cover the range of the forms predicted from the general consideration. Also, we find the situation in which the flow of new links, increasing via a power law, takes out the growing network from the class of free-scale structures.

We start from the most general considerations. 
In the scale-free networks, a wide range of the connectivity distribution function is of the power-law form, $P(q) \propto q^{-\gamma}$. It will be clear from the following that, to keep the network in the class of free-scale nets, the flow of new links has to be a power function of the number of sites of the network that plays the role of time, i.e., be proportional to $t^\alpha$. 

First, let us assume that the exponent of the distribution is less than two. The reasonable range is $1<\gamma<2$. To produce the restricted average connectivity (that is proportional to $t^\alpha$), the distribution has to have a cut-off at large $q$. Its natural value is of the order of the total number of sites, that is proportional to $t^{\alpha+1}$. It is also clear from the following that, if the distribution is non-stationary, 
the connectivity distribution of the scale-free networks has to be of the form, $t^z q^{-\gamma}$, between 
$q \sim t^x$ and $q \sim t^{\alpha+1}$, where $z$ and $x$ are some exponents.  
It is a restricted function in the range below $q \sim t^x$. $z \geq 0$ since the network grows. From the normalization condition, $\int_0^\infty dq\,P(q,t)=1$, we get $x=z/(\gamma-1)$. The average connectivity, $\overline{q}$ is of the order 
$t^{\alpha+1}/t$, then $t^\alpha \sim \int^{t^{\alpha+1}} dq\,q\,t^z q^{-\gamma} \sim 
t^{z+(2-\gamma)(\alpha+1)}$ (the value of the integral is defined by its upper limit). Therefore, $\alpha=z+(2-\gamma)(\alpha+1)$, and, finally, $\gamma= 1+(1+z)/(1+\alpha)$. One sees that $z<\alpha$ to keep the exponent $\gamma$ below $2$ as it 
was 
assumed. Also, one sees that the lower boundary for $\gamma$, $1+1/(1+\alpha)$, is approached for the stationary distribution, $z=0$. In this case, the form of the distribution is completely fixed by the accelerating growth, the exponent $\gamma$ depends only on $\alpha$. 

We do not know the values of $\alpha$ for real networks.
Obviously, $\alpha$ cannot exceed $1$ (the total number of links has to be smaller than $t^2/2$ since one may forbid multiple links). Hence, $\gamma>3/2$. 
The value $\alpha=0$ corresponds to permanent density of links \cite{dms00}. 
The density of connections in the real networks remains rather low all the time \cite{fff99}, so 
one may reasonably assume that $\alpha$ is small. Therefore, the lower boundary of the possible values of $\gamma$ is close to $2$.

The other possibility is $\gamma>2$. In this case, again $x=z/(\gamma-1)$ but the integral for the average connectivity is defined by its lower limit, 
$t^\alpha \sim \int_{t^{z(\gamma-1)}} dq\,q\,t^z q^{-\gamma} \sim t^{z-z\gamma-2)/(\gamma-1)}$. Hence, $\gamma=1+z/\alpha$, and $z>\alpha$. Thus, we have described the possible forms of the connectivity distribution. 

Let us demonstrate how these distributions may arise in the nonlinearly growing networks with the preferential linking.
We do not restrict ourselves to some model generating the increasing flow of new links dynamically in the process of the growth since it would narrow the range of possible dependencies on time for this flow. (We do not discuss here possible mechanisms of such generation. One of them was considered in \cite{v00}.) 
Instead of that, we prefer to introduce this dependence 
{\em ab initio}. 

Let us introduce the following simple model of the growing network with {\em directed} links. We consider only the distribution of incoming links, so, here, 
the connectivity is the number of incoming links $q_s$ of the site $s$.
Let each instant, a new site be added to the network.  
Let it initial connectivity be $n$, where $n\geq 0$. That means that it has $n$ incoming links from some non-specified old sites. 
Let extra $c_0 t^\alpha$ new directed links be distributed between old sites. 
We assume that each of these links comes out from an unspecified site and 
is directed to some site $s$ chosen with probability proportional to the sum of its connectivity and some constant, $q_s+A$. Here $n+A$ has to be positive. 

In fact, we generalize the model \cite{dms00} to the case of the increasing number of new links.  
The introduced power-law dependence of the number of new links keeps the network in the class of free-scale networks but changes crucially the connectivity distribution. 

Note that, unlike the first model of a scale-free growing network with preferential linking proposed by Barab\'{a}si and Albert \cite{ba99}, the constant $A$ is introduced in the rule of the distribution of the new links between the sites of the network. The sum $n+A$ characterizes the initial ``attractiveness'' of a site for new links \cite{dms00}. 

We use the continuous connectivity approach \cite{baj99,dms00,dm00} which gives exact values for the scaling exponents of such systems as it was demonstrated in \cite{dms00}. Each site is labeled by the time of its birth, $0<s\leq t$. The connectivity distribution of the site $s$ in the continuous-$q$ approximation has the form 
$p(q,s,t)=\delta(q-\overline{q}(s,t))$ \cite{dms00}, where $\delta(\ )$ is the delta-function and $\overline{q}(s,t)$ is the average connectivity of the site $s$ at time $t$. 

The resulting equation for $\overline{q}(s,t)$ is very simple:

\begin{equation}
\frac{\partial\overline{q}(s,t)}{\partial t} = c_0\, t^\alpha 
\frac{\overline{q}(s,t)+A}{\int_0^t du\, [\,\overline{q}(u,t)+A\,]}
\,   
\label{1}
\end{equation}   
with the boundary condition $\overline{q}(t,t)=n$. One may also set $\overline{q}(0,0)=0$. 
The equation describes the distribution of flow of new links between sites according the introduced rules.  
Applying $\int_0^t ds$ to both sides of Eq. (\ref{1}) one gets 
$\int_0^t du\, \overline{q}(u,t) = nt + c_0 t^{\alpha+1}/(\alpha+1) $, so 

\begin{equation}
\frac{\partial\overline{q}(s,t)}{\partial t} =  
\frac{c_0\, t^\alpha}{(n+A)t + c_0 t^{\alpha+1}/(\alpha+1) 
}\, [\,\overline{q}(s,t)+A\,]
\, .  
\label{2}
\end{equation}  

The general solution of Eq. (\ref{1}) is  

\begin{equation}
\overline{q}(s,t)+A = 
g(s) \left[ 1 + (n+A)\frac{1+\alpha}{c_0}t^{1+\alpha} 
\right]^{\!\! 1+1/\alpha} \!\!\!
t^{-(\alpha-1/\alpha)}
\! ,  
\label{3}
\end{equation} 
where $g(s)$ is an arbitrary function of $s$. 
It is fixed by the boundary condition, $\overline{q}(t,t)=n$. Thus one obtains the result for 
 $\overline{q}(s,t)$,

\begin{equation}
\frac{\overline{q}(s,t)+A}{n+A}  = 
\left[ \frac{ 1 + (n+A)(1+\alpha)t^{-\alpha}/c_0 }
{ 1 + (n+A)(1+\alpha)s^{-\alpha}/c_0 } \right]^{\!\! 1+1/\alpha} \!\!\!
\left( \frac{s}{t} \right)^{-(\alpha+1)}
\! .  
\label{4}
\end{equation} 
In the interval $[(n+A)(1+\alpha)/c_0]^{1/\alpha} \ll s \ll t$,

\begin{equation}
\overline{q}(s,t)  = (n+A)\,\left( \frac{s}{t} \right)^{-(\alpha+1)}
\, .  
\label{5}
\end{equation} 
Thus, the exponent $\beta$, $\overline{q}(s,t) \propto s^{-\beta}$, equals $1+\alpha$ and is bigger than $1$. 
The dependence $\overline{q}(s)$ becomes constant, 

\begin{equation}
\overline{q}(s,t)  = 
(n+A)^{-1/\alpha} \left( \frac{c_0}{1+\alpha} \right)^{1+1/\alpha} t^{\alpha+1} 
\, ,
\label{6}
\end{equation} 
at $s \ll [(n+A)(1+\alpha)/c_0]^{1/\alpha}$. One may compare the result, Eq. (\ref{6}), with the total number of links in the network, $N(t) \approx c_0 t^{\alpha+1}/(1+\alpha)$. 

In the continuous-$q$ approximation, one may easily find the distribution $P(q,t)$ using the obtained  
$\overline{q}(s,t)$:

\begin{eqnarray}
P(q,t) & = & \frac{1}{t} \int^t_0 ds\, p(q,s,t) =  
\frac{1}{t} \int^t_0 ds\, \delta ( q - \overline{q}(s,t) ) 
\nonumber
\\
[5pt]
& = & - \frac{1}{t} 
\left( \frac{\partial \overline{q}(s,t)}{\partial s} \right)^{-1}
\, .
\label{7}
\end{eqnarray} 
Therefore, in the region 
$1 \ll q/(n+A) \ll \left\{\, c_0/[(n+A)(1+\alpha)]\,\right\}^{1+1/\alpha}t^{1+\alpha}$,
the connectivity distribution has the following form:

\begin{equation}
P(q,t) = \frac{(n+A)^{1/(1+\alpha)}}{1+\alpha}\, q^{-[1+1/(1+\alpha)]}
\, .
\label{8}
\end{equation} 
Thus, we obtain the stationary connectivity distribution, 
$P(q) \propto q^{-\gamma}$ with $\gamma=1+1/(1+\alpha)$ that belongs to one of the types described above. 

The last result follows also 
from the relation $\beta(\gamma-1)=1$ between the scaling exponents, which may be obtained from the assumption that $\overline{q}(s,t)$ and $P(q,t)$ are power-law functions of $s$ and $q$ correspondingly \cite{dms00}. 
Note that if $A \to \infty$, the network is out of the class of scale-free networks for any $\alpha \geq 0$.

The introduction of the increasing flow of new links in the problem changes crucially the distribution of the connectivities. Indeed, in the case of a constant density of links, 
the value of the scaling exponent $\gamma$ of the connectivity distribution varies from $2$ to infinity depending on the network parameters, $n$ and $A$ \cite{dms00}, and the values of the exponent $\beta$ are between $0$ and $1$. Here, 
for 
the increasing density of links, we obtain $\gamma$ below 2 
and the exponent $\beta$ exceeds $1$. The values of the exponents are independent of $n$ and $A$. 

To demonstrate the other possibility, $\gamma>2$, we consider below the model with a different rule of the distribution of new links. 
We make the only change in the studied above model. Let 
now a new link be directed to some site $s$ with probability proportional to 
$q_s(t)/\overline{q}(t) + B$, where $q_s(t)$ is the connectivity of site $s$, 
$\overline{q}(t)$ is an average connectivity of sites of the network, and $B$ is some positive constant. 
$\overline{q}(t) \cong c_0 t^{\alpha}/(1+\alpha)$,
so the new links are distributed between sites with probability proportional to 
$q_s(t) + B c_0 t^{\alpha}/(1+\alpha)$, where $a_0$ is positive constant. 
Hence, the previously introduced $A$ becomes to be time-dependent.   

Note that, if $\alpha=0$, this rule for the preferential linking is also reduced to the known model of the growing network with a permanent input flow of new links \cite{dms00}. 

\begin{figure}
\epsfxsize=85mm
\epsffile{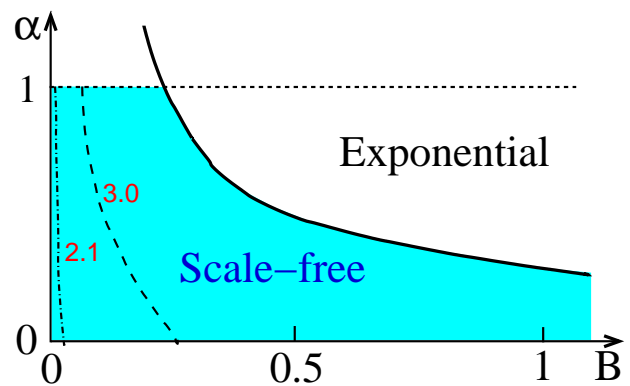}
\caption{
Phase diagram of the networks with the accelerating growth under consideration. The networks are out of the class of scale-free nets (``exponential'') above the line $\alpha = 1/B$. The exponent $\gamma$ equals $3$ on the dashed line and $2.1$ (value for the World Wide Web) on the dash-dotted one. 
$\gamma<2$ on the line $B=0$. 
}
\label{f3}
\end{figure}

Repeating the previous calculations, one gets the equation 

\begin{equation}
\frac{\partial\overline{q}(s,t)}{\partial t} = c_0\, t^\alpha 
\frac{\overline{q}(s,t) + B c_0\, t^{\alpha}/(1+\alpha)}
{nt + B c_0\, t^{\alpha+1}/(1+\alpha) + c_0 t^{\alpha+1}/(\alpha+1) } 
\label{10}
\end{equation}
with the boundary condition $\overline{q}(t,t)=n$. At long times, one obtains

\begin{equation}
\frac{\partial\overline{q}(s,t)}{\partial t} = 
\frac{1+\alpha}{1+B}\, 
\frac{\overline{q}(s,t) + B c_0\, t^{\alpha}/(1+\alpha)}{ t } 
\, .
\label{11}
\end{equation}
The solution of Eq. (\ref{11}) is 

\begin{equation}
\overline{q}(s,t) = 
\left[ n + \frac{B c_0\, s^\alpha}{1-B\alpha}
\right] 
\left( \frac{s}{t} \right)^{-(1+\alpha)/(1+B)}
- \frac{B c_0\,  t^\alpha}{1-B\alpha}
\, 
.
\label{12}
\end{equation}
If $B=0$, we obtain the previous result, $\beta=1+\alpha$. 
For $s^\alpha \gg n(1-B\alpha)/(B c_0)$, 

\begin{equation}
\overline{q}(s,t) \approx  
\frac{B c_0 t^\alpha}{1-B\alpha} 
\left\{ \left(\frac{s}{t} \right)^{\alpha-(1+\alpha)/(1+B)} - 1 \right\} 
\, .
\label{13}
\end{equation}
Therefore, the scaling exponents of the growing network are 
$\beta = (1+\alpha)/(1+B) - \alpha = (1-B\alpha)/(1+B)$ 
and 
$\gamma = 1+1/\beta = 1 + [ (1+\alpha)/(1+B) - \alpha ]^{-1} = 
2 + B(1+\alpha)/(1-B\alpha)$. The connectivity distribution 
differs sharply from the distribution obtained for the previous model. 
It is nonstationary and 
is of the form $P(q,t) \sim t^{-1+(1+\alpha)(1-B\alpha)}q^{-[1 + (1+B)/(1-B\alpha)]}$ for $q \gg t^\alpha$.
In this case, $\beta<1$ and $\gamma>2$ for any positive $\alpha$ and 
$B$. The scaling regime is realized if $B\alpha<1$. The general phase diagram for both considered model is shown in Fig. \ref{f3}.

Note 
that, in both considered cases, one cannot set $\alpha=0$ directly in 
the obtained expression for the scaling exponents. In such a situation, we get from Eqs. 
(\ref{2}) or (\ref{10}) $\beta=[1+(A+n)/c_0]^{-1}$ and $\gamma=2+(A+n)/c_0$ 
 \cite{dms00}.

It is known that the used continuous approach  
gives exact results for the scaling exponents 
of the growing networks with a constant density of connections \cite{dms00}. Nevertheless, 
it is approximate, so we have checked the obtained above results by simulation. 

The results of the simulation of considered models are shown in Figs. \ref{f1} and \ref{f2}. 
The size of networks in both studied cases is $10000$ sites. 
The number of the attempts equals 
$1000$.
In Fig. \ref{f1}, we present the log-log plots of the average connectivity vs number of a site 
for $\alpha=0.5, n=1 , A=1.0, c_0=1.0$ (the first model) and for 
$\alpha=0.5, n=1, B=0.15, c_0=1.0$ (the second one). 
In Fig. \ref{f2}, for these values of parameters of the models, we show the log-log plots of the connectivity distribution. 

The obtained values of the scaling exponents are 
within the error of the simulation from the corresponding ones found analytically. 
The values 
$\beta= 1.46 \ (1.5)$  are obtained from the simulation and analytically (in brackets) for the first model 
with the written out 
parameters,  
$\beta= 0.85 \ (0.804)$ are the corresponding values for the second model. 
$\gamma= 1.69 \ (1.667)$ and  $\gamma= 2.19 \ (2.243)$ are the values of the critical exponent of the connectivity distribution obtained for the first and for the second models, relatively. 
One may see that the correspondence is really good.     

Several different values of the scaling exponent of the distribution of incoming links in 
the World Wide Web were published (as far as we know, any data on the exponent $\beta$ are absent yet). The available data are $\gamma = 2.1$ \cite{ajb99,krrt99,bkm00} and $1.94$ \cite{ah00}. The strong difference between these values may be explained by the difference of the sizes of the scanned areas of Web. 
The most huge area was studied in \cite{bkm00}, so the value $\gamma=2.1$ seems to be the best one. 
As we have noted, one may assume reasonably that $\alpha$ is small in the real networks. We have shown that, in such a situation, the lower boundary for the possible values of $\gamma$ is slightly below $2$. We have demonstrated that, for $\gamma>2$, the connectivity distribution has to be non-stationary if the growth of the network is accelerating. 
There are no data that let us learn
whether the connectivity distributions of the World Wide Web and the Internet
are stationary or not. Our results make this question intriguing. 

The World Wide Web is still in the initial stage of its evolution. 
Perhaps, the parameters of the accelerating growth will change. In this case, our answers demonstrate the possibility of changing of $\gamma$. We have shown that it may become even less than $2$ in future.    

To demonstrate all the existing possibilities we have considered the models of growing networks with the particular rules of the preferential attachment of new links. 
These models cover the range of possibilities but provide us only with particular values of the scaling exponents. Of course, there exists a lot of additional factors (aging \cite{dm00} and dying \cite{ba00a,dm00} of sites, etc.) which may change these particular values.


In summary, we have studied the nonlinear, accelerating growth of the scale-free networks. 
We have demonstrated that it can be one of the most significant factors defining their structure. We have described the possible connectivity distributions of such networks and have fixed the lower boundary for the scaling exponent $\gamma$. 
Only the power-law time-dependence of the input flow of new links can keep the network inside of the class of scale-free networks. 
Nevertheless, we have found the region of parameters in which the scale-free structure is impossible (see the phase diagram in Fig. \ref{f3}).
Our results demonstrate possibility of quite different scenarios of the network evolution and let us hope to approach satisfactory description of the real networks.
\\

SND thanks PRAXIS XXI (Portugal) for a research grant PRAXIS XXI/BCC/16418/98. JFFM 
was partially supported by the project PRAXIS/2/2.1/FIS/299/94. We also thank A.N. Samukhin 
 for many useful discussions. 
\\
$^{\ast}$      Electronic address: sdorogov@fc.up.pt\\
$^{\dagger}$   Electronic address: jfmendes@fc.up.pt

\begin{figure}
\epsfxsize=85mm
\caption{
Log-log plot of the average connectivity of a site vs its number (birth time) for the considered models. 
For the first model, $\alpha=0.5, n=1 , A=1.0, c_0=1.0$. 
For the second model, $\alpha=0.5, n=1, B=0.15, c_0=1.0$.
The dashed lines have the slopes equal to the values of the scaling exponent $\beta$ obtained analytically.
}
\label{f1}
\end{figure}


\begin{figure}
\epsfxsize=85mm
\caption{
Log-log plot of the distribution of the number of incoming links of sites for the considered models.
For the first model, $\alpha=0.5, n=1 , A=1.0, c_0=1.0$. 
For the second model, $\alpha=0.5, n=1, B=0.15, c_0=1.0$.
The dashed lines have the slopes equal to the values of the scaling exponent $\gamma$ obtained analytically. For better presentation, the dependences are displaced along the vertical axis.
}
\label{f2}
\end{figure}

\end{multicols}


\begin{references}

\bibitem{hppj98}  B.A. Huberman, P.L.T. Pirolli, J.E. Pitkow and R.J.
Lukose, Science {\bf 280}, 95 (1998).

\bibitem{ajb99}  R. Albert, H. Jeong, and A.-L. Barab\'{a}si, Nature {\bf 401}, 
130 (1999)

\bibitem{ha99} B.A. Huberman and L.A. Adamic, Nature {\bf 401}, 131 (1999).

\bibitem{ls98}  J. Lahererre and D. Sornette, Eur. Phys. J. B {\bf 2}, 525
(1998).

\bibitem{r98}  S. Redner, Eur. Phys. J. B {\bf 4}, 131 (1998).

\bibitem{ba99}  A.-L. Barab\'{a}si and R. Albert, Science {\bf 286}, 509
(1999). 

\bibitem{ba00a} R. Albert, H. Jeong, and A.-L. Barab\'{a}si, Nature {\bf 406}, 
378 (2000); R. Albert and A.-L. Barab\'{a}si, cond-mat/0005085.

\bibitem{asbs00}  L.A.N. Amaral, A. Scala, M. Barthelemy, and H.E. Stanley,
cond-mat/0001458.

\bibitem{mn00} C. Moore and M.E.J. Newman, Phys. Rev. E {\bf 61}, 5678 (2000). 
 
\bibitem{b1} P. Baran, {\it Introduction to Distributed Communications Networks}, 
RM-3420-PR, August 1964, $<$http://www.rand.org/publications/RM/baran.list.html$>$.

\bibitem{krrt99} R. Kumar, P. Raghavan, S. Rajagopalan, and A. Tomkins, Proc. of the 25th VLDB Conference (Edinburgh, 1999), 639-650.

\bibitem{fff99} M. Faloutsos, P. Faloutsos, and C. Faloutsos, Comput. Commun. Rev. {\bf 29}, 251 (1999). 

\bibitem{bkm00} A. Broder, R. Kumar, F. Maghoul, P. Raghavan, S. Rajagopalan, R. Stata, A. Tomkins, and J. Wiener, Proc. of the 9th WWW Conference (Amsterdam, 2000), 309.  
 

\bibitem{baj99} A.-L. Barab\'{a}si, R. Albert, and H. Jeong, Physica A {\bf 272}, 173
(1999).

\bibitem{dms00} S.N. Dorogovtsev, J.F.F. Mendes, and A.N. Samukhin,  
cond-mat/0004434.

\bibitem{krl00} P.L. Krapivsky, S. Redner, and F. Leyvraz, cond-mat/0005139.


\bibitem{dm00}  S.N. Dorogovtsev and J.F.F. Mendes, 
cond-mat/\\0001419, 
Phys. Rev. E {\bf 62}, 1842 (2000); cond-mat/0005050, 
Europhys. Lett. {\bf 52}, .... (2000).



\bibitem{v00} A. V\' azquez, cond-mat/0006132.




\bibitem{ah00} L.A. Adamic and B.A. Huberman, cond-mat/0001459. 
 

\end{references}
\end{document}